\documentclass[12pt,preprint]{aastex}

\newcommand{\e}            {\mbox{$^{-1}$}}

\newcommand{\simgt}        {\gtrsim}

\newcommand{\pp}           {\noindent\hangindent 20pt\hangafter=1}
\def\n2hp{\mbox{N$_2$H$^+$}}
\def\c34s{\mbox{C$^{34}$S}}

\def\t1c{\mbox{$\theta^1$\,Ori\,C}}

\def\plotfiddle#1#2#3#4#5#6#7{\centering \leavevmode
\vbox to#2{\rule{0pt}{#2}}
\includegraphics{#1}}

\begin{document}

\title{The Masses of the Orion Proplyds from Submillimeter Dust Emission}
\author{Jonathan P. Williams and Sean M. Andrews}
\affil{Institute for Astronomy, 2680 Woodlawn Drive,
Honolulu, HI 96822; jpw@ifa.hawaii.edu, andrews@ifa.hawaii.edu}
\and
\author{David J. Wilner}
\affil{Harvard-Smithsonian Center for Astrophysics,
60 Garden Street, Cambridge, MA 02138; dwilner@cfa.harvard.edu}

\slugcomment{To appear in the Astrophysical Journal}
\shorttitle{Masses of the Orion proplyds}
\shortauthors{Williams et al.}

\begin{abstract}
We have imaged the 880 $\mu$m continuum emission from the
``proplyds'' in the center of the Trapezium Cluster in Orion
using the Submillimeter Array with a beam size $1\farcs 5$ FWHM and
an rms of 2.7~mJy. Five sources are detected with fluxes
in the range 18 to 38~mJy, which includes dust emission from
four proplyds and ionized gas from $\theta^1$\,Ori\,G.
The total masses of the detected proplyds derived from their dust
emission range from 1.3 to $2.4 \times 10^{-2}~M_\odot$
assuming a dust temperature of 20~K and mass opacity of 0.03~cm$^2$~g\e.
The eighteen other proplyds within the field-of-view were not formally
detected individually, but by combining the fluxes measured at their
locations, an average flux of 1.1mJy was determined for them on a
statistical basis, corresponding to a mass of $8\times 10^{-4}M_\odot$.
The four detected proplyds have sufficient disk mass bound to their
central stars to form planetary systems on the scale of our Solar System.
\end{abstract}

~

\section{Introduction}

High mass stars, on account of their luminosities, temperatures and
explosive ends, have a disproportionally large effect on the
interstellar medium relative to their numbers. Their influence appears
to extend down to the scales of star-forming cores and even planet-forming
disks. Star formation may be triggered by supernovae shock
waves in neighboring clouds (Elmegreen \& Lada 1977)
but, at closer quarters, it may be stunted by photoionization and
evaporation of molecular clumps within HII regions (Hester et al. 1996).
The dramatic {\it Hubble Space Telescope (HST)} images of
photoevaporating protoplanetary disks, or proplyds, in the Orion
Nebula beginning with O'Dell, Wen \& Hu (1993) indicate that planet
formation may, too, be inhibited. Given that most low-mass stars
are born in OB associations (McKee \& Williams 1997), it is
important to study disk evolution in such harsh environments.

The Trapezium Cluster in Orion is the nearest young, massive star
forming region and it is consequently the most intensively studied
(e.g. O'Dell 2001). Approximately 1200 young stars with ages
$\sim 1$~Myr have been identified toward the central 1~pc diameter
of the cluster core (McCaughrean et al. 2002).
A single star of spectral type O6ep, \t1c, dominates the ultraviolet
radiation field and ionizes the gas cocoons around nearby low-mass stars.
The ionized gas is unbound and evaporates from the stars.
Mass loss rates averaging $\sim 10^{-7}~M_\odot$~yr\e\
were first inferred from centimeter wavelength observations by
Churchwell et al. (1987).
The aforementioned HST data showed that many of the ionized
knots are tadpole-shaped structures with a tail extending away
from \t1c\ (O'Dell et al. 1993), and near-infrared imaging revealed
low-mass stars at their centers (McCaughrean \& Stauffer 1994).
The visibility of the central stars despite the inferred large
mass of circumstellar material led all of these authors to suggest
that the ionized gas originates from the surfaces of externally
photoionized protoplanetary disks.

As the integrated mass loss of a proplyd over the age of the cluster
is larger than typical disk masses around T Tauri stars,
it was immediately apparent that their evolution was largely being
dictated by their environment.
The disk mass that remained, however, could not be determined from
the observations. A lower limit can be obtained for the those objects
that are seen in silhouette against the bright nebular background but
are sufficiently far enough away from \t1c\ that they are not strongly
ionized and do not produce significant optical emission
(O'Dell \& Wen 1994; McCaughrean \& O'Dell 1996; Bally et al. 2000;
Smith et al. 2005).
Based on its resolved size and a minimum column density to block the
background light, McCaughrean et al. (1998) determined a lower limit
to the mass of the giant proplyd 114-426 of
$5\times 10^{-4}~M_\odot \simeq 0.5~M_{Jupiter}$.
This is twenty times lower than the $0.01~M_\odot$ minimum formation
mass of our Solar System (Weidenschilling 1977) and thus this
technique was not able to yield useful constraints on
whether the proplyds are viable candidates for planet
formation on scales of our own Solar System.

Disk masses are best measured
at longer wavelengths where the dust emission becomes optically thin.
Interferometry is essential to resolve the tightly clustered proplyds
from each other and also to filter out the strong emission from the
background molecular cloud.
Mundy, Looney, \& Lada (1995) used the BIMA interferometer at
$\lambda 3.5$~mm to image a field around \t1c\ containing 33 proplyds.
Several significant peaks were found, four coincident with proplyds,
but the intensity was consistent with free-free emission from
ionized gas and they were unable to measure masses.
By analyzing the non-detections, however, they were able to place
a statistical upper limit of $0.03~M_\odot$ on the average disk mass.
The dust emission increases at shorter wavelengths and the free-free
emission decreases. Using the OVRO array, Bally et al. (1998a)
imaged two fields containing a total of six proplyds at $\lambda 1.3$~mm,
made a tentative detection of one object
and placed upper limits of $0.015~M_\odot$ on the other objects.
Lada (1999) presented a mosaic of two fields at $\lambda 1.3$~mm
with the Plateau de Bure interferometer that claimed three detections.
The implied masses were $\sim 0.01~M_\odot$ but these have not been
analyzed in detail.

The Submillimeter Array\footnotemark\footnotetext{
The Submillimeter Array is a joint project between the Smithsonian
Astrophysical Observatory and the Academia Sinica Institute of Astronomy
and Astrophysics, and is funded by the Smithsonian Institution and the 
Academia Sinica.} (SMA; Ho et al. 2004) is ideally suited to a
renewed look at this issue. Located on Mauna Kea, it can operate at
shorter wavelengths than other interferometers and is therefore very
sensitive to dust emission. Furthermore, it has a relatively large field
of view which allows many proplyds to be imaged simultaneously. 
In this paper, we present SMA observations at $880~\mu$m of a
single $32''$ full width half maximum (FWHM) field of view
toward 23 proplyds around \t1c.
We detect five sources and measure the average flux for the rest.
The observations are described in \S2, our results are presented in
\S3, and we discuss the implications for planet formation in \S4.

\section{Observations}
A single field containing 23 proplyds around \t1c\ was
imaged with the SMA using six antennae in its compact configuration
on November 13 and 14, 2004.
The phase center of the observations was
$\alpha(2000)=5^{\rm h} 35^{\rm m} 16^{\rm s}.6$, 
$\delta(2000)=-5^\circ 23' 28\farcs 1$, slightly south-west of \t1c\
to maximize the number of proplyds in the field.
The receivers were tuned in double sideband mode to place the
CO(3--2) line in the center of the upper sideband.
Each sideband provided the correlator with a 2~GHz spectral region
separated from the central frequency of 340.7~GHz ($880~\mu$m)
by $\pm 5$~GHz.
Weather conditions were very good, and
system temperatures varied from 300 to 700~K over
different baselines as the field moved from $30^\circ$
to $65^\circ$ elevation.

The amplitude and phase were monitored by alternately
interleaving 5-minute integrations of the quasars J0423-013
and J0609-157 with each 20 minute observation on source. The stronger
J0423-013 was used to perform the final calibration, and J0609-157
was imaged as a check on the system performance.
The bandpass response and flux scaling were determined from
a 30 minute observation of Uranus toward the end of each track.

The data for each night were calibrated with the MIR software package.
The atmosphere was very stable during both nights and the rms phase
error after calibration was $\sim 10^\circ$ on short baselines
increasing to $\sim 20^\circ$ on the longer baselines.
The passband, gain, and flux calibrated data from each night
were then combined during the imaging process in MIRIAD.
The total integration time on the \t1c\ field was 9.6 hours.
Strong CO emission was detected and the continuum data were
extracted by summing over the line free channels.

The shortest projected baseline was $b_{min}=16~{\rm k}\lambda$
and the observations are therefore largely insensitive to features
with angular scales greater than $\lambda/2b_{min}\simeq 13''$
(Wilner \& Welch 1994). This greatly reduces the potential confusion
of extended emission from the background molecular cloud
(Johnstone \& Bally 1999).

Inversion of the continuum data to create a ``dirty map''
showed a strong source near the edge of the field of view at
the $\sim 20\%$ gain of the primary beam, coincident with the
proplyd 159--350 using the nomenclature of O'Dell \& Wen (1994).
This source produced artifacts in the southeast part of the map
that proved difficult to clean,
and we therefore removed a point source at its location
from the visibilities before inverting and cleaning to analyze
the other sources in the field.
The cleanest map and lowest rms in the central region was obtained by
removing a 45~mJy source but the source properties derived so far outside
the primary beam half power point are highly uncertain.
Natural weighting of the $uv$-data resulted in a beamsize
of $1\farcs 9\times 1\farcs 2$ at position angle $303^\circ$ east
of north and the cleaned map that is presented here was restored with a
circular $1\farcs 5$ beam that has the same area as the naturally
weighted beam. The rms noise noise level is $\sigma=2.7$~mJy~beam\e.

\section{Results}
Contours of the continuum emission are shown in Figure~1.
The lefthand image shows the location of the four Trapezium O stars
and optically identified proplyds relative to the data.
The righthand image shows the contours overlaid on the HST
image from Bally et al. (1998b) and shows the morphologies,
brightness and colors of the proplyds.
Five proplyds were detected within the $32''$ FWHM of the primary beam
with a peak flux greater than $3\sigma = 8.1$~mJy~beam\e\
and are marked with a red cross in the lefthand image of Figure~1.
The two brightest sources, 163-317 and 170-337, were seen in
subsets of the data, both lower and upper sidebands, and
both the first and second day of observations.

Integrated fluxes for the five sources were measured by summing the pixels
within the $3\sigma$ contour level and correcting for primary
beam attenuation. The emission from 170-337 and 171-340 are
merged and the flux for each one was estimated by cutting
across the $3\sigma$ contour by eye.
Most of the proplyds in the field of view were detected at
centimeter wavelengths with the VLA
(Churchwell et al. 1987; Felli et al. 1993)
and have spectral energy distributions that are consistent with
thermal bremsstrahlung (Garay, Moran, \& Reid 1987).
None of the sources that we have detected are variable at 3.6~cm
(Zapata et al. 2004), suggesting that any contribution
from gyrosynchrotron emission is small and that we are not seeing
a bright flare.

Thermal bremsstrahlung emission decreases gradually toward
shorter wavelengths but dust emission sharply increases.
Spectral energy distributions (SEDs) for the five SMA detections
are plotted in Figure~2. A combination of bremsstrahlung and
modified ($\beta=1$) blackbody emission are overlaid. These fits
are poorly constrained due to the small number of data points
where the SED slopes are changing. Nevertheless, they are adequate
to show that the $880~\mu$m flux is dominated by dust emission
in all but one case, 167-317,
and to estimate the contribution from ionized gas emission in
the other four. The measured SMA fluxes, corrected for primary
beam attenuation, the peak signal-to-noise ratio, and
estimated contribution from bremsstrahlung emission are
listed in Table~1.

167-317 ($\theta^1$\,Ori\,G) is the brightest proplyd at
centimeter wavelengths and in the Mundy et al. 3.4~mm map.
The ionized gas emission is strong even at $880~\mu$m
and our SED fit shows that the dust emission contributes
only 4.6~mJy to the measured flux. There is some scatter in
the centimeter fluxes and a 20\% uncertainty in the millimeter
fluxes, however, so this number is uncertain.
Nevertheless, it does not appear to be a significant detection of
dust emission.
For the other four sources, however, there is a clear dust excess
which allow us to measure the proplyd masses for the first time.

Masses were calculated from the dust excess flux at $880~\mu$m
in the standard way way using a distance to Orion of 450~pc and the
same Hildebrand (1983) formulation for the grain opacity
$\kappa_\nu = 0.1(\nu/1200~{\rm GHz})$
as in Mundy et al. (1995) and Bally et al. (1998a).
Implicit in this value is a gas-to-dust ratio of 100
and assumptions about the grain properties.
These are discussed further in \S4.
We assumed a lower temperature, 20~K, than previous work
based on the average for disks in Taurus-Auriga (Andrews \& Williams 2005)
and because the bulk of the disk dust mass lies in the midplane
and is shielded from UV-heating even in the intense radiation field
at the center of the Trapezium Cluster (Chiang \& Goldreich 1997).
Masses would be lower by a factor of 3.3 if the dust temperature were 50~K.

Eighteen proplyds are undetected within the primary beam FWHM.
As with the previous interferometer observations
of the proplyds (Mundy et al. 1995; Bally et al. 1998a), we can study
the statistics of the non-detections to constrain their collective properties.
The histogram of fluxes in beam-sized apertures around these 18 proplyds
is compared to the background in Figure~2. For this comparison, a region
around each of the five detections was blanked out.
The flux distribution toward the proplyds is offset from that 
of the background and has a mean of 0.8~mJy~beam\e.
This is only a marginally significant result, however,
as the map is oversampled. The expected noise in the combination
of 18 positions is $2.7/\sqrt{18}=0.6$~mJy~beam\e\
and the histogram mean is only slightly greater than this.
Nevertheless, taken at face value and applying a correction for the
mean gain, 0.73, of the interferometer toward the 18 positions,
the intrinsic mean flux of the non-detections is 1.1~mJy, which
corresponds to a mass of $8\times 10^{-4}~M_\odot$.

The integrated CO map showed several resolved clumps of emission
but they did not appear coherent in velocity or centered on
any of the proplyds. The CO line is optically thick and traces the
gas temperature. The temperature probably varies more than the column
density in the background cloud and the interferometer map is
likely picking up hot spots in the molecular cloud-HII region
interface rather than the proplyds themselves.

\section{Discussion}
These observations have proven more successful at measuring the masses
of the Orion proplyds than previous interferometer work because we
were able to observe at shorter wavelengths and at higher sensitivity.
Our $3\sigma = 8.1$~mJy~beam\e\ detection limit corresponds to a mass
limit of $5.6\times 10^{-3}~M_\odot$ for individual sources. Allowing
for the different dust temperature assumptions, this is an order of magnitude
more sensitive than Bally et al. (1998a) who were in turn an order of
magnitude more sensitive than Mundy et al. (1995).
The IRAM Plateau de Bure map shown in Lada (1999) covers many of
the same objects and includes detections of 170-337, 171-340,
and 158-327, though the latter detection is not confirmed in our map.

The 23 proplyds within the primary beam FWHM of our map are listed
along with their morphological type, projected distance from \t1c,
associated stellar mass, and disk radius in Table~2.
The morphology and association with a microjet are taken from
the tables in Bally et al. (1998b, 2000). Distances and disk radii
are also measured from the HST images in Bally et al. (1998b).
Stellar masses are estimated from K-band magnitudes and placement
on 1~Myr evolutionary tracks by McCaughrean \& Stauffer (1994).

The statistics are meager, but the four dust detections
do not have a preferred morphological type and lie at a
range of projected distances. The stellar masses are somewhat
uncertain in absolute terms though probably good relative measures
and suggest that the detected disks tend to be associated with
the more massive ($\simgt 1~M_\odot$) stars.

The conversion from submillimeter flux to disk mass contains many
uncertainties in the grain properties and gas-to-dust ratio that
are encapsulated in the parameter $\kappa$.
The Hildebrand (1983) formulation that we have used assumes an
average grain size of $0.1~\mu$m, appropriate to interstellar clouds
and is normalized to a gas-to-dust ratio of 100.
Throop et al. (2001) and Shuping et al. (2003)
show that the typical grain sizes in the outer parts of
the giant silhouette proplyd, 114-426, are $\sim 2-5~\mu$m.
If the grains are spherical, this does not greatly change
the value of $\kappa$ at 1~mm (Pollack et al. 1994).
However, if the grains have aggregated in a fractal manner,
the cross-section per unit grain mass may be as much as an order of
magnitude higher (Wright 1987)
and our proplyd mass estimate may be substantially overestimated.
On the other hand, masses would be underestimated if there is
dust settling to the disk midplane and grain growth to
centimeter sizes and larger (Wilner et al. 2005)
or if the inner disk is optically thick (Andrews \& Williams 2005).

The gas-to-dust ratio is also highly uncertain. 
Dust settling to the disk midplane and photoevaporation of the
upper disk envelope will tend to decrease the
gas-to-dust ratio below the canonical ISM value of 100
(Throop \& Bally 2005) leading to an overestimate of the mass
using the Hildebrand $\kappa$.
We did not clearly detect CO emission from any of the proplyds
but future observations of other gas tracers that do not deplete
and are abundant in the cold, dense disk midplane
will help reveal the gas content and dynamics of the disks.

Given the uncertainty in the mass calculations, it is instructive
to compare the submillimeter fluxes from the proplyds with isolated
disks in Taurus. After correcting for the greater distance to Orion,
the proplyd flux distribution is found to more closely match Taurus
Class II than Class I objects (Andrews \& Williams 2005).
As proplyds lack outer envelopes, this is also their closest
morphological match. In this sense, and only for the four dust
detections, the Trapezium Cluster environment appears not to have
had a substantial effect on these systems.

The stars within the proplyds have ages $\sim 1$~Myr (Hillenbrand 1997)
but the evaporation timescales, equal to the disk masses measured here
divided by the mass loss rates measured by Churchwell et al. (1987),
are of order $10^{-2}~M_\odot/10^{-7}~M_\odot~{\rm yr}^{-1} \sim 10^5$~yr.
Note that if disks were originally more massive, they would
have been larger in size and their mass loss rates consequently greater.
Thus, the present disk mass divided by the present mass loss rate
is a good measure of the proplyd lifetime (St\"orzer \& Hollenbach 1999).
The resulting discrepancy in timescales may be due to a younger age,
$\sim 10^5$~yr, for the ionizing source, \t1c, than the rest of the
cluster (Johnstone, Hollenbach, \& Bally 1998)
or a lower average mass loss rate if either the ionized gas
recombines before leaving the system (O'Dell 1998) or if the
proplyds move on radial, rather than circular, orbits through
the cluster (St\"orzer \& Hollenbach 1999).

In this regard, it is intriguing to note that three of the four dust
detections were at a signal-to-noise ratio greater than 7,
and one at a signal-to-noise ratio of 4.
A smooth distribution of disk masses with increasing numbers at
smaller masses should have resulted in proportionally
more detections at lower signal-to-noise.
The number of detections is small but hints at an abrupt mass loss
event such as might be expected for close encounters to \t1c\ on
plunging orbits through the cluster. Observations of additional
proplyds at a range of distances from \t1c\ will show the disk mass
distribution and the effect of environment more clearly. If a distance
dependent effect is seen, it would suggest that most stars
move on approximately circular orbits and the influence on
\t1c\ on planet formation is limited.
If the detection rate does not increase with distance, however, it
may be that the orbits are eccentric and
that \t1c\ would then influence large numbers of potential planetary
systems as they pass through the center of the cluster.
The effect may be purely destructive through almost complete
photoevaporation of a disk but it may also promote the formation of
planets via preferential removal of the gas which allows dust disk
instabilities to develop (Throop \& Bally 2005).

The four dust detections have disk masses slightly greater than
the $0.01~M_\odot$ minimum mass solar nebula (Weidenschilling 1977).
The mass loss is concentrated in the outer parts of the disks where
the gravitational potential of the central star is weakest and
photoevaporation is most effective (Hollenbach, Yorke, \& Johnstone 2000).
The radius of the bound inner region depends on the stellar mass and
whether the gas is ionized by EUV photons or remains neutral and
only heated by the FUV radiation field. The detected proplyds lie
far enough away from \t1c\ for the second condition to apply and
the central $\sim 20-50$~AU radius of the disks survive if the central
stellar mass is $1~M_\odot$ (Johnstone et al. 1998).
Disk radii were measured to be roughly 40~AU in 170--337 and 171--340
(Bally et al. 1998b). Thus, at most, the outer 50\% of their disks
will be lost and for a surface density $\Sigma\sim r^{-3/2}$,
the surviving mass fraction is at least $\sim 60$\%.
The other two objects probably have smaller disk radii in which case
even more of the mass will survive. We conclude that for these
four proplyds, at least, the submillimeter emission indicates there
is sufficient material bound to the stars to form Solar System scale
planetary systems.

There are intriguing possibilities for the non-detections too.
We found a statisically positive flux toward the 18 proplyds
that lay below our $3\sigma$ detection limit.
The corresponding (gas + dust) mass limit is $8\times 10^{-4}~M_\odot$
and the dust-only mass limit is therefore
$8\times 10^{-6}~M_\odot\simeq 3~M_\oplus$.
That is, even if all the gas were lost from these disks,
there would still be enough material to form terrestrial-like
planets. Far more sensitive observations would be required to verify
this on an individual basis, of course, and it would also be
important to average the VLA data in a similar way to measure the
low level bremsstrahlung emission.


\acknowledgments
We thank the referee, Mark McCaughrean, for his thorough review which
helped improve the paper, and the SMA staff for carrying out the observations.
We enjoyed interesting discussions with Dave Hollenbach, John Bally,
and Michael Liu, and provocative comments from Frank Shu and Christine
Wilson concerning the collective properties of the non-detections.
Support for this work was provided by NSF grant AST-0324328 (JPW)
and NASA Origins of Solar Systems Program Grant NAG5-11777 (DJW).

\section{References}

\pp Andrews, S.~M., \& Williams, J.~P.\ 2005, \apj, submitted

\pp Bally, J., Testi, L., Sargent, A., \& Carlstrom, J.\ 1998a, \aj, 116, 854
 
\pp Bally, J., Sutherland, R.~S., Devine, D., \& Johnstone, D.\
    1998b, \aj, 116, 293

\pp Bally, J., O'Dell, C.~R., \& McCaughrean, M.~J.\ 2000, \aj, 119, 2919
 
\pp Chiang, E.~I., \& Goldreich, P.\ 1997, \apj, 490, 368

\pp Churchwell, E., Wood, D.~O.~S., Felli, M.,
    \& Massi, M.\ 1987, \apj, 321, 516

\pp Elmegreen, B.~G., \& Lada, C.~J.\ 1977, \apj, 214, 725

\pp Felli, M., Churchwell, E., Wilson, T.~L., \& Taylor, G.~B.\
    1993, \aaps, 98, 137
 
\pp Garay, G., Moran, J.~M., \& Reid, M.~J.\ 1987, \apj, 314, 535
 
\pp Hester, J.~J., et al.\ 1996, \aj, 111, 2349
 
\pp Hildebrand, R.~H.\ 1983, \qjras, 24, 267
 
\pp Hillenbrand, L.~A.\ 1997, \aj, 113, 1733
 
\pp Ho, P.~T.~P., Moran, J.~M., \& Lo, K.~Y.\ 2004, \apjl, 616, L1
 
\pp Hollenbach, D.~J., Yorke, H.~W., \& Johnstone, D.\ 2000,
    Protostars and Planets IV, 401

\pp Johnstone, D., Hollenbach, D., \& Bally, J.\ 1998, \apj, 499, 758 

\pp Johnstone, D., \& Bally, J.\ 1999, \apjl, 510, L49
 

\pp Lada, E.~A. 1999, NATO ASIC Proc.~540:
    The Origin of Stars and Planetary Systems, 441
 
\pp McCaughrean, M.~J., et al.\ 1998, \apjl, 492, L157 

\pp McCaughrean, M.~J., Zinnecker, H., Andersen, M., Meeus, G.,
    \& Lodieu, N.\ 2002, The Messenger, 109, 28 

\pp McCaughrean, M.~J., \& Stauffer, J.~R.\ 1994, \aj, 108, 1382
 
\pp McCaughrean, M.~J., \& O'Dell, C.~R.\ 1996, \aj, 111, 1977
 
\pp McKee, C.~F., \& Williams, J.~P.\ 1997, \apj, 476, 144
 
\pp Mundy, L.~G., Looney, L.~W., \& Lada, E.~A.\ 1995, \apjl, 452, L137
 
\pp O'Dell, C.~R., \& Wen, Z.\ 1994, \apj, 436, 194
 
\pp O'Dell, C.~R., Wen, Z., \& Hu, X.\ 1993, \apj, 410, 696
 
\pp O'Dell, C.~R.\ 1998, \aj, 115, 263
 
\pp Pollack, J.~B., Hollenbach, D., Beckwith, S., Simonelli, D.~P.,
    Roush, T., \& Fong, W.\ 1994, \apj, 421, 615
 
\pp Shuping, R.~Y., Bally, J., Morris, M., \& Throop, H.\ 2003,
    \apjl, 587, L109
 

\pp Smith, N., Bally, J., Licht, D., \& Walawender, J.\ 2005, \aj, 129, 382
 
\pp St\"orzer, H., \& Hollenbach, D.\ 1999, \apj, 515, 669

\pp Throop, H.~B., Bally,~J., Esposito, L.~W., \& McCaughrean, M.~J.\ 2001,
    Science, 292, 1686 

\pp Throop, H.~B., \& Bally, J.\ 2005, \apj, 623, L149

\pp Weidenschilling, S.~J.\ 1977, \apss, 51, 153
 
\pp Wilner, D.~J., \& Welch, W.~J.\ 1994, \apj, 427, 898

\pp Wilner, D.~J., D'Alessio, P., Calvet, N., Claussen, M.~J., \& Hartmann, L.
    2005, ApJ, in press (astro-ph/0506644)
 
\pp Wright, E.~L.\ 1987, \apj, 320, 818
 
\pp Zapata, L.~A., Rodr{\'{\i}}guez, L.~F., Kurtz, S.~E., \& O'Dell, C.~R.\
    2004, \aj, 127, 2252

\clearpage
\begin{figure}[ht]
\plotfiddle{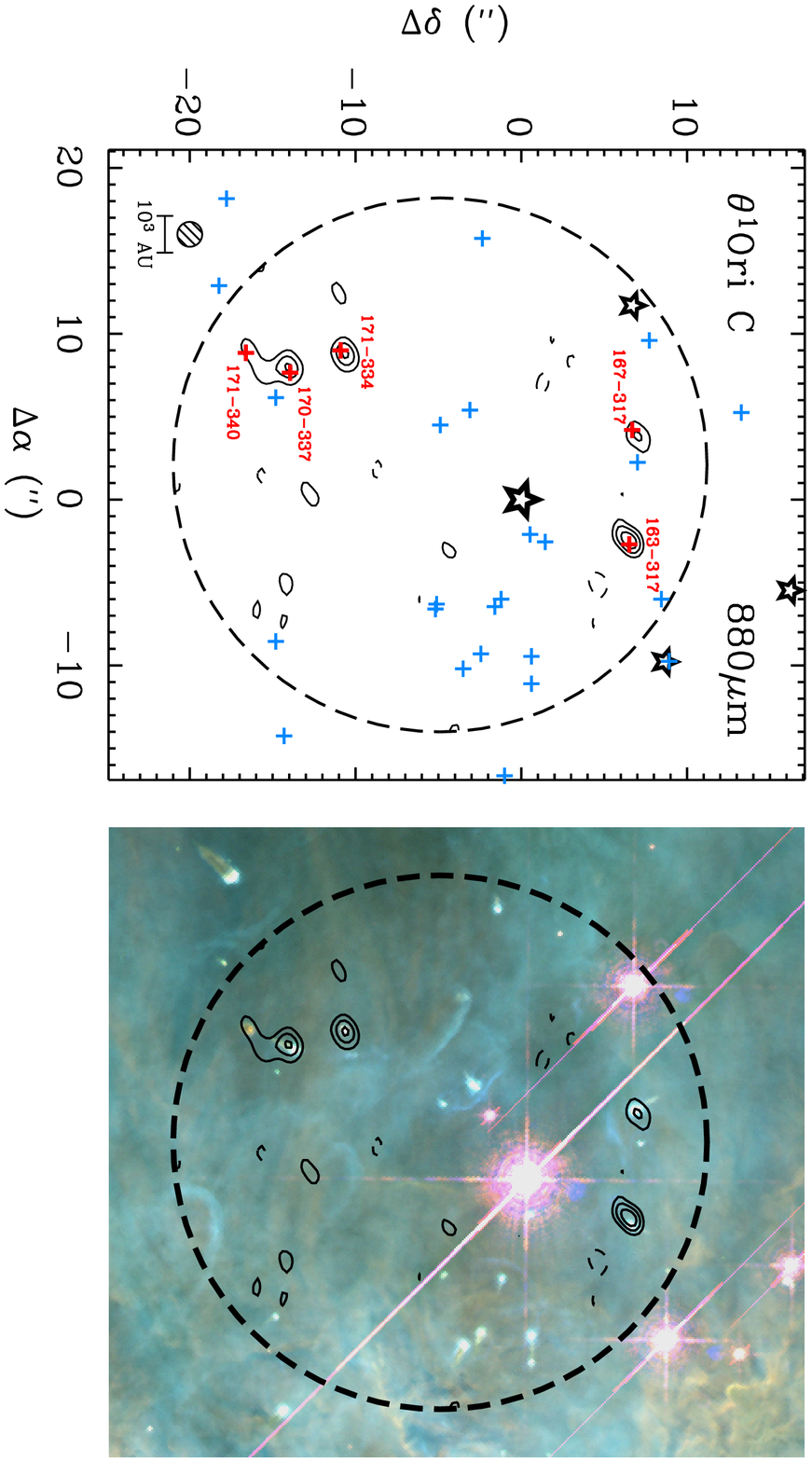}{0pt}{90}{60}{60}{240}{-300}
\end{figure}
\vskip 3.3in
\noindent{\bf Figure 1:}
Contours of $880~\mu$m continuum emission toward the proplyds
in the Trapezium Cluster. The lefthand image shows the locations
of the proplyds; red crosses for the 5 detections labeled
with the O'Dell \& Wen nomenclature and blue crosses for the
18 non-detections. The position of the four Trapezium O stars
are shown by the large star symbols and the center of the
coordinate grid has been set to \t1c.
The $1\farcs 5$ synthesized beam and scale bar are shown in
the lower left corner.
The contour levels are $3,5,7\times\sigma$ where $\sigma=2.7$~mJy~beam\e\
is the rms noise level in the map.
The righthand image shows the same contours overlaid on the
HST image from Bally et al. (1998b).
In both images, the large dashed circle shows the FWHM of the primary beam,

\clearpage
\begin{figure}[ht]
\plotfiddle{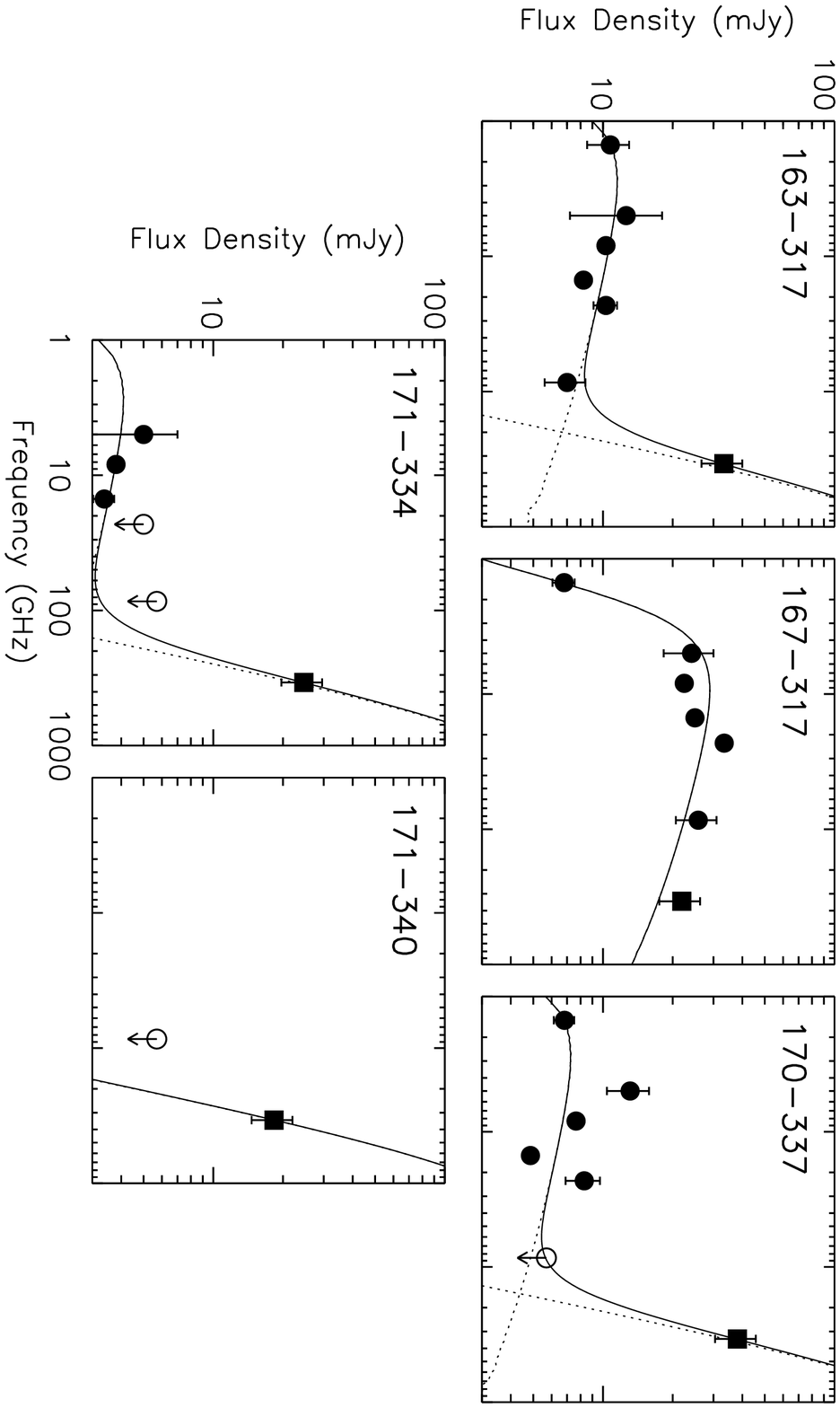}{50pt}{90}{65}{65}{265}{-300}
\end{figure}
\vskip 3.4in
\noindent{\bf Figure 2:}
Spectral energy distributions from 20~cm to $880~\mu$m
for the five SMA detections. The fluxes shortward of 30~GHz
are from VLA observations by Felli et al. (1993),
Garay et al. (1987) and Zapata et al. (1994).
In the case of multi-epoch observations, the error
bar shows the range of measured values.
The 86~GHz fluxes are from BIMA observations by Mundy et al.
and is estimated for 163-317 from the $3\sigma$ contour in their map.
Upper limits are shown by open symbols.
The squares show the 341~GHz SMA measurements.
The uncertainties in the BIMA and SMA measurementss
are dominated by uncertainties in the flux scales, estimated to 20\%.
Fits to the ionized gas and dust emission are overlaid.

\clearpage
\begin{figure}[ht]
\plotfiddle{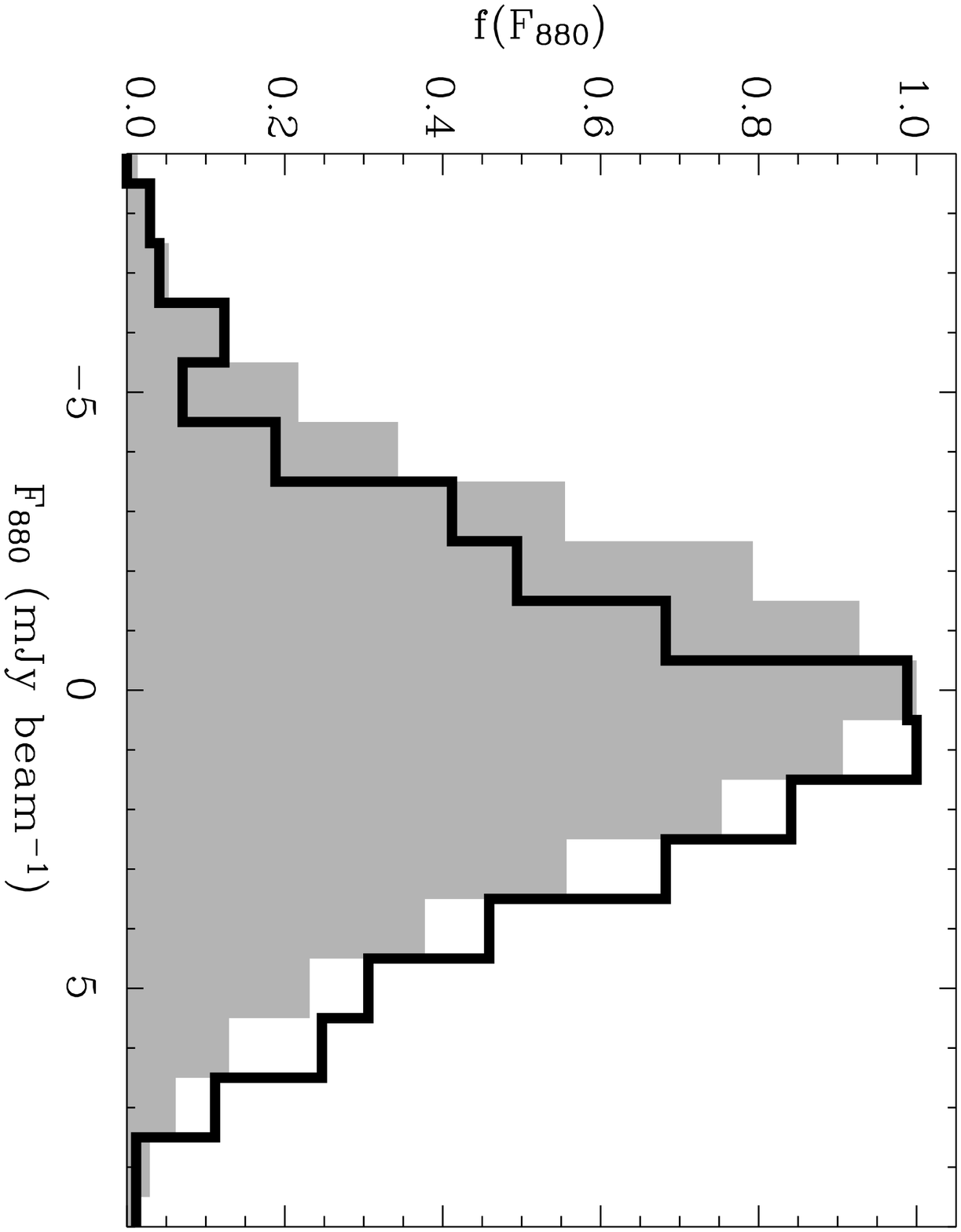}{50pt}{90}{55}{55}{220}{-280}
\end{figure}
\vskip 3.7in
\noindent{\bf Figure 3:}
Normalized histograms of the flux distribution toward the
undetected proplyds (line) and the rest of the map within
the primary beam FWHM (grayscale).
The five detected proplyds have not been included in either plot
and the fluxes were not corrected for primary beam attenuation.
The flux distribution toward the undetected proplyd positions is offset
from the background flux distribution and has a mean value 0.8~mJy~beam\e.

\clearpage
\begin{table}
\begin{center}
TABLE 1\\
Masses of the detected proplyds\\
\vskip 2mm
\begin{tabular}{lcccc}
\hline\\[-2mm]
Proplyd & $F_{880_{\mu{\rm m}}}^{\rm a}$ & peak S/N & Bremsstrahlung$^{\rm b}$ & $M^{\rm c}$ \\
        & (mJy) &  & (mJy) &($10^{-2}M_\odot$)\\[2mm]
\hline\hline\\[-3mm]
163--317           &  33.3  &  8.7  &    5.9    &   1.9   \\
167--317$^{\rm d}$ &  21.9  &  5.3  &   17.3    & \nodata \\
170--337           &  38.1  &  7.3  &    3.7    &   2.4   \\
171--334           &  24.6  &  7.5  &    2.1    &   1.6   \\
171--340           &  18.3  &  4.0  &    0      &   1.3   \\[3mm]
\hline\\[-3mm]
\multicolumn{5}{l}{$^{\rm a}$ Observed flux, corrected for primary beam attenuation}\\
\multicolumn{5}{l}{$^{\rm b}$ Estimated contribution from ionized gas}\\
\multicolumn{5}{l}{$^{\rm c}$ $\kappa=0.03$~cm$^2$~g\e, $T=20$~K}\\
\multicolumn{5}{l}{$^{\rm d}$ $\theta^1$\,Ori\,G; no significant dust excess}
\end{tabular}
\end{center}
\end{table}

\clearpage
\begin{table}
\begin{center}
TABLE 2\\
Proplyd properties\\
\vskip 2mm
\begin{tabular}{lccccc}
\hline\\[-2mm]
Proplyd & Dust excess? & Type$^{\rm a}$ & $D_*$ & $M_*$ & $R_D$ \\
        &       &      & ($10^{16}$ cm) & ($M_\odot$) & (AU) \\[2mm]
\hline\hline\\[-3mm]
163-317 & $\checkmark$ &        &     4.7   &   1.8  &       \\
170-337 & $\checkmark$ & jet,od &    10.9   &   2.2  &   40  \\
171-334 & $\checkmark$ &        &     9.5   & $>2.5$ &       \\
171-340 & $\checkmark$ & sd/e   &    12.9   &        &   40  \\[2mm]
\hline\\[-2mm]
157-323 &   $\times$   &        &     7.4   &   2.1  &       \\
158-323 &   $\times$   &        &     6.3   & $>2.5$ &       \\
158-326 &   $\times$   & sd/e   &     6.5   &   0.7  &   40  \\
158-327 &   $\times$   & sd/e   &     7.2   &        &   20  \\
159-338 &   $\times$   &        &    11.9   &   0.2  & $<20$ \\
160-328 &   $\times$   & od     &    11.1   &        &   45  \\
161-314 &   $\times$   &        &     5.1   &        &       \\
161-322 &   $\times$   &        &     4.1   &        &       \\
161-324 &   $\times$   &        &     4.2   & $<0.1$ &       \\
161-328 &   $\times$   & sd/e   &     5.1   &   0.1  &   40  \\
163-322 &   $\times$   &        &     2.0   &        &       \\
163-323 &   $\times$   &        &     1.4   &   0.2  &       \\
166-316 &   $\times$   &        &     4.9   &   0.8  & $<20$ \\
167-317 &   $\times$   & jet    &     5.2   &   2.4  &       \\
168-326 &   $\times$   &        &     4.2   &   0.1  &       \\
168-328 &   $\times$   &        &     4.5   &   0.7  &       \\
169-338 &   $\times$   &        &    11.1   &        &       \\
171-315 &   $\times$   &        &     8.2   &        &       \\
176-325 &   $\times$   &        &    11.0   &   0.9  &       \\[3mm]
\hline\\[-3mm]
\multicolumn{6}{l}{$^{\rm a}$ sd/e = silhouette disk in bright envelope}\\
\multicolumn{6}{l}{$^{\rm ~}$ od=disk seen in O[I] emission}
\end{tabular}
\end{center}
\end{table}

\end{document}